\documentclass[twocolumn]{aa}
\usepackage{graphicx}
\usepackage{txfonts}

\def\sec{\ifmmode {}^{\prime\prime}\else ${}^{\prime\prime}$\fi~}
\def\secdot{\hbox{$.\!\!^{\prime\prime}$}}  
\def\mindot{\hbox{$.\!\!^{\prime}$}}
\def\magdot{\ifmmode {}^{\rm m}\!\!\!.\, \else ${}^{\rm m}\!\!\!.\,$\fi}
\def\daydot{\ifmmode {}^{\rm d}\!\!\!.\, \else ${}^{\rm d}\!\!\!.\,$\fi}
\def\asec{\ifmmode ^{\prime\prime}\else$^{\prime\prime}$\fi}

\begin{document}

\authorrunning{Fabrika et al.}
\title{Crowded field 3D spectroscopy of LBV candidates in M\,33}

\author{S.\,Fabrika \inst{1}, O.\,Sholukhova \inst{1},
T.\,Becker \inst{2}, V.\,Afanasiev \inst{1}, M.\,Roth \inst{2}, 
S.F.\,Sanchez \inst{2}}

\institute{
Special Astrophysical Observatory, Nizhnij Arkhyz,
Karachaevo-Cherkesia, Russia 369167\\
\email{fabrika@sao.ru}
\and
Astrophysikalisches Institut Potsdam,
An der Sternwarte 16, D-14482, Potsdam, Germany\\
\email{mmroth@aip.de}
}

\date{}

\abstract{
We present integral field spectroscopy of the LBV 
candidate stars B\,416 and v\,532 in the local group galaxy M\,33. 
B\,416 is surrounded by an elongated  ring--like nebula, which has 
a projected radius of $20 \times 30$~\,pc. 
From the datacube we create ionization and radial velocity maps of 
the nebula.  The excitation of the gas decreases towards the outer part
of the ring, while the inner part of the nebula is filled with a more excited 
gas. In the EW direction the ring is seen to expand with a maximum
projected velocity amplitude of about 40\,km/s. The eastern 
part approaches the observer. We estimate the nebula dynamical lifetime 
$\sim 8 \cdot 10^5$~years. It could be a residual MS bubble,
which indicates a main--sequence or pre--LBV status of the star.
We classify B\,416 as an ``iron star'' or B[e]--supergiant. 
In v\,532 an elongated nebula has been marginally detected.
The total projected size of the nebula along the main axis is 30~pc, 
and the total radial velocity gradient is $44 \pm 11$~km/s.
v\,532 exhibits both strong photometric and spectral variability. 
At the time of our observations it was in an intermediate
brightness state with a rich nitrogen spectrum. 
We classify v\,532 as an LBV, showing LBV\,$\leftrightarrow$\,Ofpe/WN 
transitions. 
We stress the importance of 
integral field spectroscopy as the optimal technique for studying nebulae 
and the evolution of LBV--like stars in nearby galaxies. 
}

\titlerunning{Crowded Field 3D Spectroscopy of LBV candidates}

\maketitle


\section*{Introduction}

Analysis of the brightest supergiants as individual stars in external 
galaxies provides a unique tool for determining the properties 
of young stellar populations in their host galaxy.
In particular, nearby galaxies are ideally suited to studying the most massive
stars systematically, both during their normal evolution in different 
environments, as well as during the short and unstable stages of evolution 
as OB supergiants, rare hypergiants, Luminous Blue Variables (LBVs),
Ofpe/WN stars, B[e]-supergiants (Kudritzki~1998; Lamers et al.~2001),
or during their latest stages of evolution as peculiar massive binary objects 
with  relativistic jets and accretion disks like SS\,433 
(Fabrika and Sholukhova~1995). All of these stars are young and massive, 
and they are among the brightest stars in their host galaxy. 
They share some main spectral properties like a blue continuum
and strong emission lines, and they often show a surrounding nebula.
In the Milky Way a considerable fraction of such stars may
be hidden by dust extinction in the galactic plane, which is why they are
observed best in nearby galaxies.

The evolutionary connections between the various classes
of massive stars, such as the OB-stars, B[e]-supergiants, LBVs,
Ofpe/WNL, and Wolf-Rayet stars
are not well understood. Systematic surveys of LBV--like candidates
in massive galaxies (Calzetti et al.~1995, Massey et al.~1996, Corral 1996,
Fabrika and Sholukhova~1999) provide a basis for isolating these 
objects for subsequent follow--up and quantative spectroscopy (Massey et
al.~1996, Sholukhova et al.~1997, 1999) to
reveal the basic properties of these objects. 
In M\,33 there are 4 confirmed LBV stars and about a dozen LBV candidates,
but only one known Ofpe/WN star, and about 5 new candidates for this 
latter class of objects (Massey et al.~1996).
Not a single SS433-type star has been found in M\,33, and also no 
B[e]-supergiant has been confirmed to date. 

LBVs are extremely massive stars ($M > 40 \, M_{\sun}$) at an unstable stage
of evolution with a mass loss rate of $\sim 10^{-4}\, M_{\sun}/y$. 
They show strong
spectral and photometrical variability on time-scales from months to years
(Humphreys and Davidson~1994; Lamers et al.~1998a).
LBVs represent an instability stage after the main sequence. 
It is not clear whether there is a relation between LBVs 
and B[e]-supergiants (Conti~1976), since their spectra often look practically
the same, but the latter stars do not show strong variability
(Zickgraf et al.~1986; Lamers et al.~1998b). 
Spectral properties of B[e]-supergiants
can be interpreted by a non--isotropical mass--loss occurring in the 
equatorial plane (Zickgraf et al.~1985; Zickgraf et al.~1986).
They are probably a subclass of rapidly rotating post main-sequence
stars. 
One may expect an evolutionary connection between the B[e]--supergiants 
and the LBVs.
 
Important information can be derived from morphological studies
of the nebulae surrounding such stars. Because of their heavy mass loss, 
LBV stars are expected to show extended envelopes, which consist 
of ejected matter and swept-up interstellar material. 
The same is true for B[e] supergiant stars. Direct imaging and kinematic  
studies help to shed light on the mass loss history and on the physics of the
stellar wind. 

Numerical models of the nebular expansion around massive stars 
(Garcia-Segura et al., 1996) have shown that those nebulae reach diameters 
of up to 20 - 40 pc at the main-sequence and pre-LBV stages. 
Such scales correspond to $\sim 10$\sec at the distance of M\,33
(3\secdot5/pc). LBV nebulae are smaller because of a shorter evolution
time. Practically all known LBV stars have circumstellar shells
(Humphreys and Davidson~1994). 
Typical LBV nebula sizes are in the range of 0.1~-- 4~pc, expansion
velocities 15~-- 100 km/s, and dynamical times are in the range
100~-- $5 \cdot 10^4$~years (Nota~1999; Figer et al.~1999; Weis~2003). 
There are possibly larger gas nebulae around LBV stars, for example
a ring--like nebula around S\,Dor has a size up to 70~pc (Weis~2003). 

Spectroscopic
analysis of the star is often hampered by the presence of strong nebular
emission lines which contaminate the stellar spectrum,
and is complicated in crowded fields.
For our targets we choose the technique of 
{\em Integral Field Spectroscopy} (IFS, sometimes also called 
{\em 3D Spectroscopy}) in order to optimize the separation 
of the stars from their crowded environments. It provides both 
improved subtraction of
nebular contamination from the stellar spectrum and
spatially resolved nebular spectra. 
One can make extensive use of the full
2-dimensional information contained in 3D data, which can be also
described as stacks of very narrow bandwidth filter images (datacubes).

Here we report on results of IFS observations of two LBV candidates in the
Local Group galaxy M\,33, the stars B\,416 and v\,532. The first star was 
designated as blue star N\,416 in the photographic survey by Humphreys and 
Sandage~(1980). It was also identified as an LBV candidate in follow--up
spectroscopy of the UV--brightest stars of M\,33 (Massey et al.~1996).
In a program searching for SS\,433--like stars in M\,33,
Shemmer and Leibowitz (1998) detected periodic microvariations in the
photometry of B\,416. These oscillations were confirmed as periodic and 
coherent by Shemmer et al.~(2000), who classified the star as LBV based
on the spectral similarity in comparison with typical LBV spectra at a 
quiescence stage. The coherent brightness oscillations may become quite 
important for our understanding of the physics of LBV stars; however, B\,416 
does not show the large amplitude variability typical for LBVs. 

The second object of this paper is the variable blue star v\,532 (Romano 1978; 
Artyukhina et al.~1995), which is known to exhibit the characteristic 
variability 
of LBVs, i.e. an amplitude of $\approx1^m$ on a time--scale of years 
(Kurtev et al.~2001). Short-scale and quasi-periodical variabilities have 
also been found in this star (Sholukhova et al., 2002). Strong
emission lines, spectral and photometrical variability (Szeifert 1996;
Fabrika 2000; Sholukhova et al. 2002; Polcaro et al. 2003) indicate  
that v\,532 is an LBV object.

\section*{Observations and data reduction}

\begin{figure*}
   \centering

\includegraphics[width=13cm]{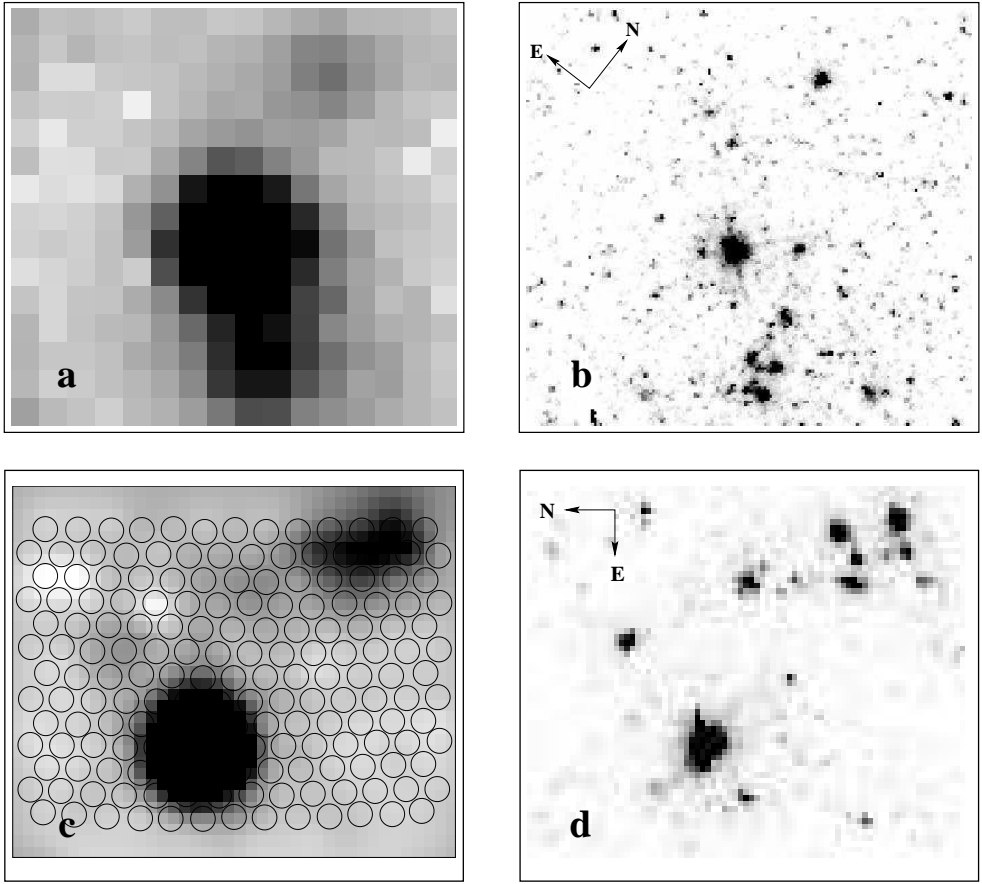}
\vspace{1cm}
 \caption{
 {\bf(a)} MPFS map of B416 at 6000~{\AA} (spatial sampling 1~arcsec,
 the FOV is 16$\times$15~arcsec$^2$) and the same field
 {\bf(b)} from the HST WFC2 image.
 An enlarged fraction of this field from the INTEGRAL in the same wavelength
 {\bf(c)} and the corresponding enlarged HST field {\bf(d)}.
 Note the different orientation of the MPFS and INTEGRAL fields.
 }
\end{figure*}

Observations of v\,532 and B\,416 were carried out with the Multi-Pupil
Fiber Spectrograph 
MPFS\footnote{http://www.sao.ru/\~{}gafan/devices/mpfs/mpfs\_main.htm}
(Afanasiev 1998) at the prime focus of the 6m telescope BTA (Russia). 
The integral field unit (IFU) of 16$\times$\,15 square spatial 
elements (''spaxels'')
covers  a rectangular region of 16\sec$ \times$\,15\sec 
on the sky (1\sec per element).
Optical fibers  transmit the light from the 240 spaxels of the IFU 
together with 16 additional fibers located at a radius of 
$\approx 4\mindot5$ outside of the IFU for the purpose of sampling the 
sky background far away from the field. 
In nearby galaxies like
in M33 the sky fibers will still be located within the galaxy, thus 
we did not use the sky fibers for the background subtraction.
The entire set of 256 fibers is reformatted to form
a pseudo-slit, whose emerging light is dispersed and 
projected onto the focal plane
of the spectrograph. We used a SITe TK1024 backside-illuminated CCD with
1024$\times$1024 pixels and a pixel size of 24$\mu$m. 

Two 20-minute exposures of v\,532 were taken on September 18, 1998 with
a reciprocal dispersion of 1.3~\AA/pixel (FWHM~$\approx 3.5$\,\AA) 
in the spectral
range 4470~-- 5800~\AA\AA\ with a seeing of 1.5--2~arcsec FWHM. 
B\,416 was observed on September 28, 1998 in two 20-minute exposures with
a reciprocal dispersion of 2.6~\AA/pixel (FWHM~$\approx 7$\,\AA) 
in the spectral
range of 4250~-- 6900~\AA\AA\ with a seeing of 3~arcsec FWHM.

3D observations of B\,416 were also carried out with 
INTEGRAL\footnote{http://www.ing.iac.es/\~{}bgarcia/integral/html/integral\_home.html}
at the 4.2m WHT, La Palma (Arribas et al.\ 1998) on January 18, 2001.
This instrument is a bare fiber bundle type of 3D spectrograph.
Contrary to MPFS, the IFU has no lensarray in front of the fibers, resulting in
a hexagonal package with small gaps between the circular fiber apertures.
We used the SB1 bundle, which has 205 fibers and covers a
rectangular  FOV of 7.8$\times$6.4~arcsec$^2$ 
with a projected fiber diameter of 
0\secdot45. The fiber bundle is coupled to the bench-mounted fiber 
spectrograph WYFFOS on the Nasmyth platform of the WHT. 
The detector is a thinned,
backside-illuminated 1K$\times$1K, 24~$\mu$m pixel CCD. We used an 
R300B grating, 
giving a spectral coverage of 5800~{\AA} and a spectral resolution of 11~{\AA} at 
a reciprocal dispersion of 5.8~{\AA}/pixel and took 2 exposures of 1800 sec
each. The conditions were clear with a seeing of 0.9--1.0\sec FWHM.

Data reduction for both the MPFS and INTEGRAL data sets was performed with the
P3d package (Becker 2002), which was originally developed for the PMAS 
instrument 
(Roth et al. 2000), but has also been shown to work well for other instruments.
The raw CCD frames were bias and dark subtracted, and cleaned from cosmic ray 
events. 
The spectra extraction was performed using a profile fitting
algorithm, which simultaneously solves for cross-talk between adjacent
spectra and straylight patterns on the detector.
P3d incorporates a robust algorithm to reliably determine the geometry of
the spectra.
After extraction a wavelength calibration was applied using arc exposures. The
wavelength-dependent fiber-to-fiber transmission variation was calibrated 
by means of twilight flatfield exposures.
At this final stage of data reduction a 2-dimensional
image of stacked spectra appears. 
Monochromatic maps or an entire {\em datacube} may be subsequently derived 
from this stacked spectra format. 
The final analysis of the fully reduced 3D data, i.e.\ generation of 
monochromatic maps at selected wavelengths, co-adding flux within 
a digital aperture, sky subtraction,
etc., was performed with the MONOLOOK tool  (Becker 2002). 

The full 2-dimensional spatial information allows one to
accurately model the background of resolved or unresolved stars and 
gaseous emission at any wavelength. 
A detailed description of this
technique is given in Roth et al.\ (2003). 
Application of image deconvolution techniques is able to 
disentangle the different components from severely crowded stellar 
fields with significant contamination from nebular emission.
We used the {\it cplucy} two-channel algorithm, which is available in IRAF. 
A more detailed account of this technique is given in
Becker (2002) and Becker et al.\ (2003).

In Fig.\,1 we present continuum maps of B\,416 from a wavelength
region near 6000~\AA\ of MPFS and INTEGRAL, respectively.
For each IFU map the same FOV as observed with HST\footnote{Based on 
observations made with the NASA/ESA Hubble Space Telescope, obtained 
from the data archive at the Space Telescope Science Institute. 
STScI is operated by the Association of Universities
for Research in Astronomy, Inc. under NASA contract NAS 5-26555.}
is shown for comparison.  While the MPFS map (a)
represents the intensity distribution exactly over the square spaxels
of the lens array, the INTEGRAL map (c) was interpolated from 
the 205 data points of the fiber bundle (overlay). Note the patches
of low intensity near the upper left corner in (c) indicating the presence
of dead fibers in the bundle.

%

\begin{figure*}
\centering
\includegraphics[width=17cm]{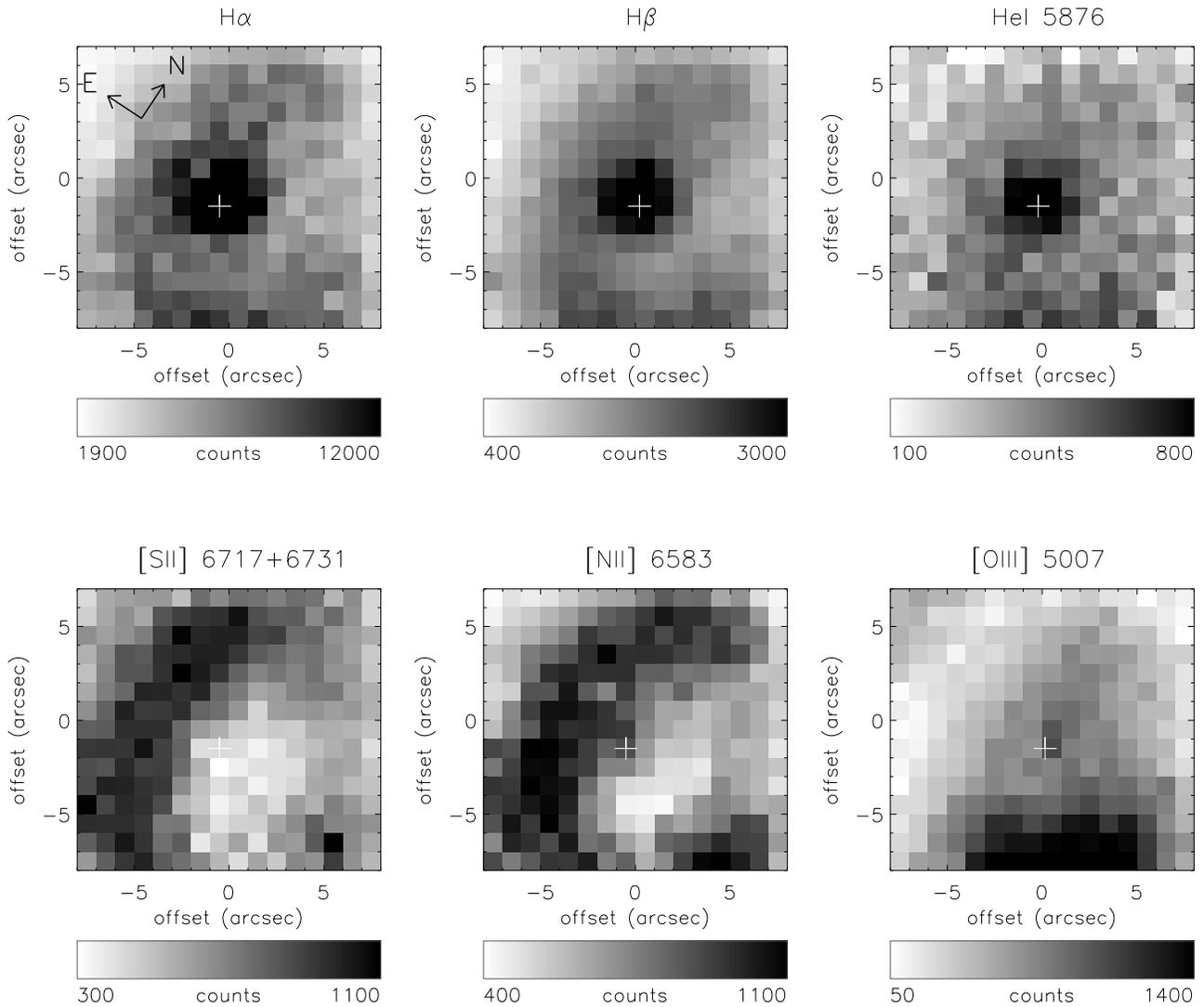}
\caption{
Monochromatic MPFS maps of B\,416 taken in central 7.5~\AA\ bands of
different emission lines,
continuum subtracted. Intervals of relative fluxes (counts) used in the
maps are shown.
The position of the star is indicated by a white cross. A prominent
arc-like nebular feature is visible to the left. The star itself
shows emission in the Balmer lines and He\,I.
}
\end{figure*}


While the MPFS observations suffered from rather poor seeing,
the conditions with INTEGRAL were more favourable and yielded much better
spatial resolution. 
In Fig.\,1(a) the image of B\,416 is blended with the star cluster
located from 2\secdot5 to 5\secdot5 to the SW. 
In Fig.\,1(c) the cluster is 
clearly separated from B\,416 and marginally resolved 
into stars. Three other isolated stars seen in the HST frame to the 
W of B\,416 are also marginally present in the UNTEGRAL map.

\section*{Results}

\subsection*{The nebula around B\,416}

Figure\,2 presents monochromatic images of B\,416 in the emission lines of
H$\alpha$, H$\beta$, He\,I~$\lambda5876$, [SII]\,$\lambda 6717+\lambda6731$,
[NII]~$\lambda 6583$, and [OIII]\,$\lambda 5007$ from the MPFS observations.
The images were produced in $\pm 2.5$\,\AA\ bands from the central pixels of
the lines (7.5\,\AA\ bands) as relative fluxes. Corresponding continuum 
images were substracted. 
There is a shift in the monochomatic pictures depending on wavelength due
to atmospheric dispersion. Its value is 0\secdot42/1000\,\AA\
(0\secdot65 between H$\alpha$ and [OIII]~$\lambda 5007$) towards the NW 
with decreasing wavelength.
The central star position is indicated in the figure by crosses in each image
taking the atmospheric shifts into account.


A ring--like nebula is clearly seen in the emission line maps of Fig.\,2. 
It is non-symmetrical and presents a different morphology in lines of different 
excitation. The star itself is a source of emission in permitted lines. 
The stellar emission in H$\alpha$ is particularly strong and gives rise
to a high level of contrast, which is why the nebula appears faint
at this wavelength. There is no apparent forbidden line emission coincident
with the PSF of the central star. The inner parts of the nebula are filled
with faint and diffuse emission in [OIII]~$\lambda 4959, 5007$. 

In order to disentangle nebula and stellar PSF components in the
spectral domain we attempted a gaussian decomposition of the nebular
and stellar emission line profiles in H$\alpha$ and H$\beta$. Since the
nebular line is intrinsically narrow, but the stellar emission line profile 
is expected to be broader under the influence of a strong stellar wind, this 
approach seemed to provide a useful criterion.

As a best guess we estimated the width of the nebular emission in H$\alpha$
and [NII]~$\lambda 6548, 6584$ in regions far away from the star, yielding a
gaussian FWHM\,$\approx 6.8$~\AA, which is in fact identical to the spectral
resolution. We then fitted the observed spectrum near H$\alpha$ by three
gaussians for H$\alpha$ and [NII]~$\lambda 6548, 6584$, whose widths 
are equal to that of 
the instrumental profile, and by a fourth component for the stellar 
H$\alpha$ emission, whose FWHM was determined from the fit.
In fact, for all spaxels close to the centroid of the star, this latter 
component makes a non-negligible contribution to the spectrum and has an
FWHM\,$\approx 12.2$~\AA. The broad component contributes to 55\,\% of the
total H$\alpha$ emission inside of the centroid of the star, but it
is absent in spaxels far away from the star.
The observed spectrum in a spaxel close to the star, 
the three narrow and one broad gaussian fits, 
and the sum of the fits are plotted in Fig.\,3.
This spectrum was extracted in a spaxel not far from the star to show both
broad and narrow components of H$\alpha$. The flux units are the same as in
Fig.\,2, whose images were produced in 7.5\,\AA\ bands.
Similarily, the decomposition of H$\beta$ into broad and narrow profiles 
resulted in a broad line width of FWHM\,$\approx 10$~\AA. This value agrees 
well with the value found in H$\alpha$ and with the corresponding intrinsic 
velocities. 

\begin{figure}[h]
\includegraphics[width=6cm, angle=-90]{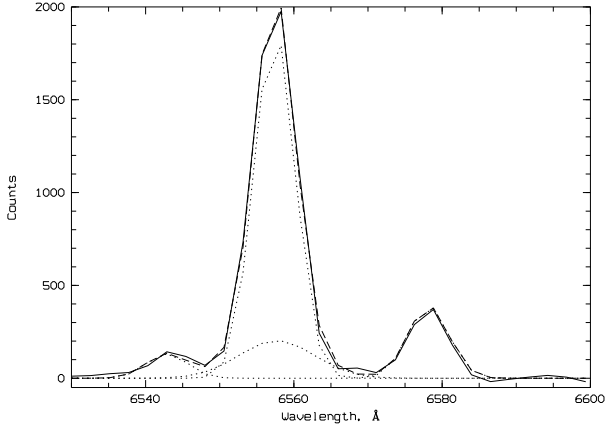}
\caption{
An example of the H$\alpha$ and [NII]~$\lambda 6548, 6584$ lines
(solid line) in a region close to the star. They are
fitted with four gaussian profiles (dashed lines)~---
broad and narrow H$\alpha$ lines and two narrow [NII] lines. A sum of the
gaussian profiles is shown by a long--dashed line
}
\end{figure}

\begin{figure*}
\includegraphics[width=17cm]{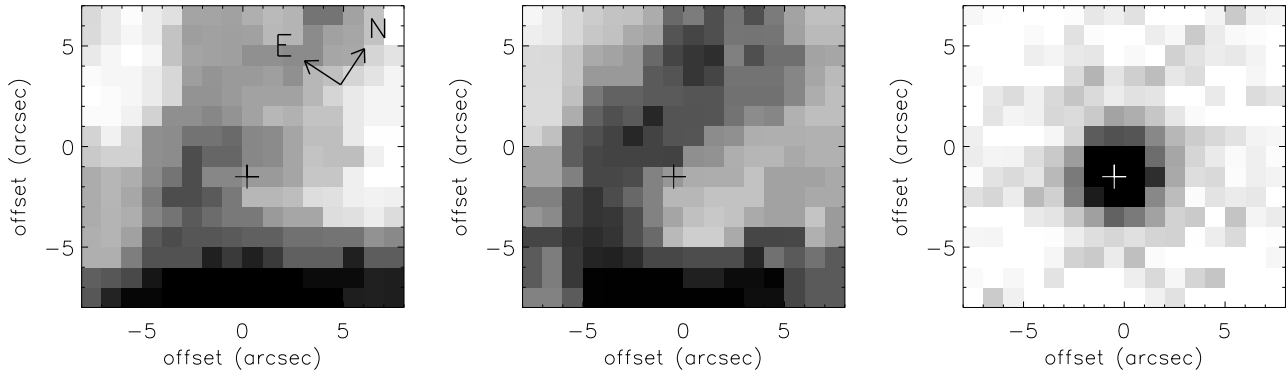}
\caption{
  Maps of B416: narrow H$\beta$ component (left panel), 
  narrow H$\alpha$ (middle), and broad H$\alpha$ (right).
  The stellar centroid is indicated by a cross in each frame.
  }
\end{figure*}

Figure\,4 demonstrates a {\em spatial} representation of the 
spectral decomposition into narrow and broad profiles, where the broad 
component is associated with the star and is perfectly coincident with the
stellar PSF. In the nebular analysis, which
follows below, we only use the narrow Balmer line component.
These results suggest that the broad Balmer emission lines are indeed most
likely formed in the atmosphere of the star. We must stress both that
an expanding
nebula near the star (with an angular extent below our spatial resolution)  
cannot entirely be ruled out, and that even the stellar 
atmosphere can contribute to the narrow emission.

The morphology of the ring--like nebula presents an almost perfect elliptical
ring (Fig.\,2), which is however interrupted on its NW side by a distinct gap.
The overall surface brightness distributions in H$\alpha$, [NII], and [SII] are 
fairly similar, but show differences in the exact size and morphological
details. The surface brightness distribution in [OIII] nebula is strikingly
different from those in the H$\alpha$, [SII], and [NII] lines.
The extended faint emission in the N, W parts and very bright SW regions 
are dominated by high excitation gas contrary to those parts of opposite 
directions. The H$\beta$ nebula pattern (and H$\gamma$, which is not shown here)
is somewhere intermediate between those of high excitation ([OIII]) and
low excitation ([SII] and [NII]).  

To illustrate the whole nebula around B\,416, we present in Fig.\,5 a small
(roughly 1~arcmin) subfield from H$\alpha$ and [OIII] images of M33 taken by
Massey et al. (2001) in the course of an imaging survey of 
Local Group galaxies, though the H$\alpha$ nebula was first detected by Shemmer et al. (2000).
The images of Massey et al. (2001) were obtained with the CTIO and 
KPNO 4-m telescopes using the Mosaic CCD cameras. 
Narrow--band filters were used 
with bandwidths of 55~\AA~ in [OIII] 
and 80~\AA~ in H$\alpha$.
The MPFS field and corresponding emission line isophotes of the narrow 
H$\alpha$ line and [OIII]~$\lambda 5007$ are overplotted.
Each step in the isophotes corresponds to 200 counts, the same relative units 
as used in Fig.\,3. The maximum isophote level is 1900 counts in H$\alpha$ and
900 counts in [OIII]~$\lambda 5007$. The MPFS maps are in excellent agreement 
with these direct images. 

The [OIII] emission around B\,416 is faint. It is concentrated on the
SW side of the nebula, where an extended [OIII] region does exist,
which is only partly covered by the MPFS field. The most distant SW part 
of the nebula (just outside of the MPFS map) has about the same shape in 
[OIII] and H$\alpha$ (Fig.\,5). This fact, together with the specific 
shape of the [OIII] nebula, may point towards an explanation that the most 
distant SW part of the whole nebula is shaped not 
only under the influence of B\,416, but also by a cluster of hot stars. 
Two compact clusters of stars (Fig.\,1b) are located in these regions,
coinciding with the [OIII] bright emission of the nebula, where, conversely,
[SII] and [NII] are relatively faint. Yet another high excitation knot is
seen within the diffuse [OIII] region, 
just 3\sec SW of the edge of the MPFS FOV
(Fig.\,5). One could speculate that the bright [OIII] emission is due to
photoionization from hot stars in these clusters, whose relation, however,
to the ring-like nebula is not immediately clear.

A similar situation is observed in the nebula DEM~L106 around LMC
B[e]--supergiant Hen~S22 (Chu et al. 2003). The H\,II region N30B with 
a cluster of hot stars is located
inside the large nebula ($\sim 30$~pc in radius).
A bow-shock-like halo around N30B is oriented towards Hen~S22. It is formed 
as a result of interaction with the stellar wind of Hen~S22. 
The halo reflects the supergiant radiation. One could conclude that the most 
distant SW region of the ring--like nebula of B\,416 is probably affected 
in a similar fashion in the environment of the hot star cluster.

\begin{figure*}
\includegraphics[width=17cm,height=6.5cm]{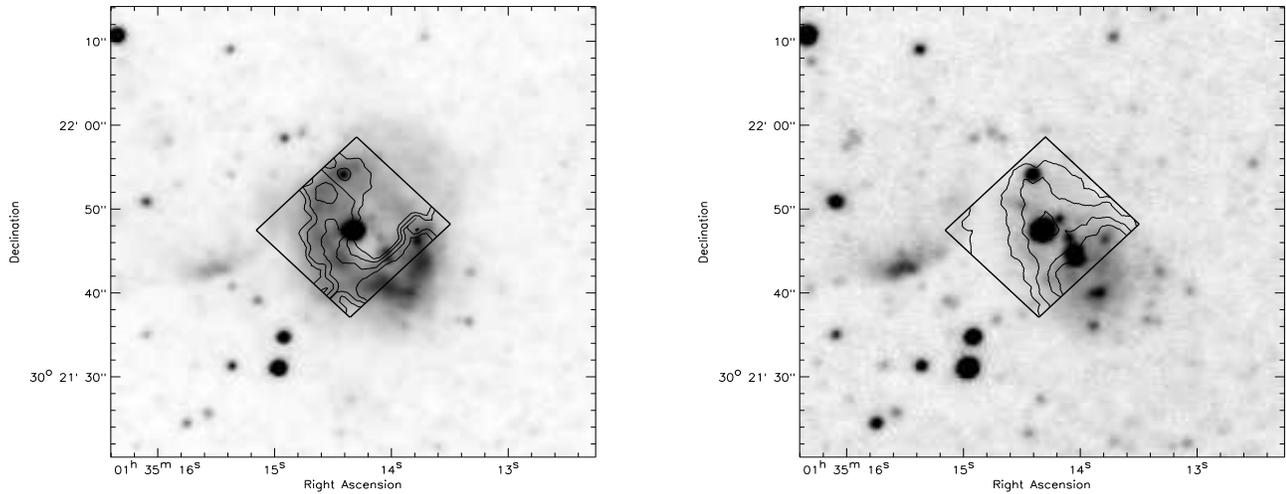}
\caption{H$\alpha$ (left) and [OIII]~$\lambda 5007$ (right) images of the 
B\,416 region taken by Massey et al. (2001) with the MPFS field and the line 
izophotes imposed on the images. 
}
\end{figure*}

For a qualitative analysis of the excitation structure of the nebula we 
present ``ionization'' maps of B\,416 in Fig.\,6. Low excitation regions
are traced by the [SII]/H$\alpha$ map, whereas high excitation regions
appear in the [OIII]/H$\beta$ ratio map. In both cases we used the Balmer
line maps derived from the narrow line components as described above. 
One can see from a comparison of Fig.\,6, 2, and 4 that over a wide range of 
position angles, the inner parts of the ring--like nebula are dominated by
high excitation, as opposed to the outer rim of the ring, which is most
pronounced in the low ionization radiation of [SII]. This observation
is qualitatively in line with the expectation of finding (i) photoionization 
predominantly near the star and (ii) shock excitation from the interaction of 
the expanding shell with the interstellar medium at the edge of the nebula.

We measured the diagnostic line ratios of H$\alpha$/[NII], H$\alpha$/[SII]
(sum the doublets) and I(6717)/I(6731) for [SII], and plotted the 
results in the diagnostic diagrams of Sabbadin \& D'Odorico (1976).  
Our mean values for the nebula around B\,416 are as follows:
H$\alpha$/[NII]$= 3.24 \pm 0.39$, H$\alpha$/[SII]$= 2.21 \pm 0.25$, and
I(6717)/I(6731)$= 1.61 \pm 0.11$, where the r.m.s. values are for 
individual resolution elements. The mean line ratios are quite 
typical for HII regions; however, the ratios are not constant accross 
the nebula. It is directly seen
in Fig.\,2 that the rim is more distant in [SII] than in [NII] and
H$\alpha$. The outer parts of the ring, at least in the NES directions,
consist of a gas of lower excitation. This conclusion is 
confirmed by the [SII]/H$\alpha$ ratio map (Fig.\,6). The ratio is
greater than 0.4 in the outer parts of the ring, which indicates a 
collisional excitation of gas as was confirmed by Smith et al. (1993)
for SNRs in M33.
The electron density (Osterbrock 1989) over the whole nebula was 
found to be at the low density limit, n$_e~<~100\,cm^{-3}$, with an 
average value of $n_e \sim 10\,cm^{-3}$.

\begin{figure*}
\centering
\includegraphics[width=13cm]{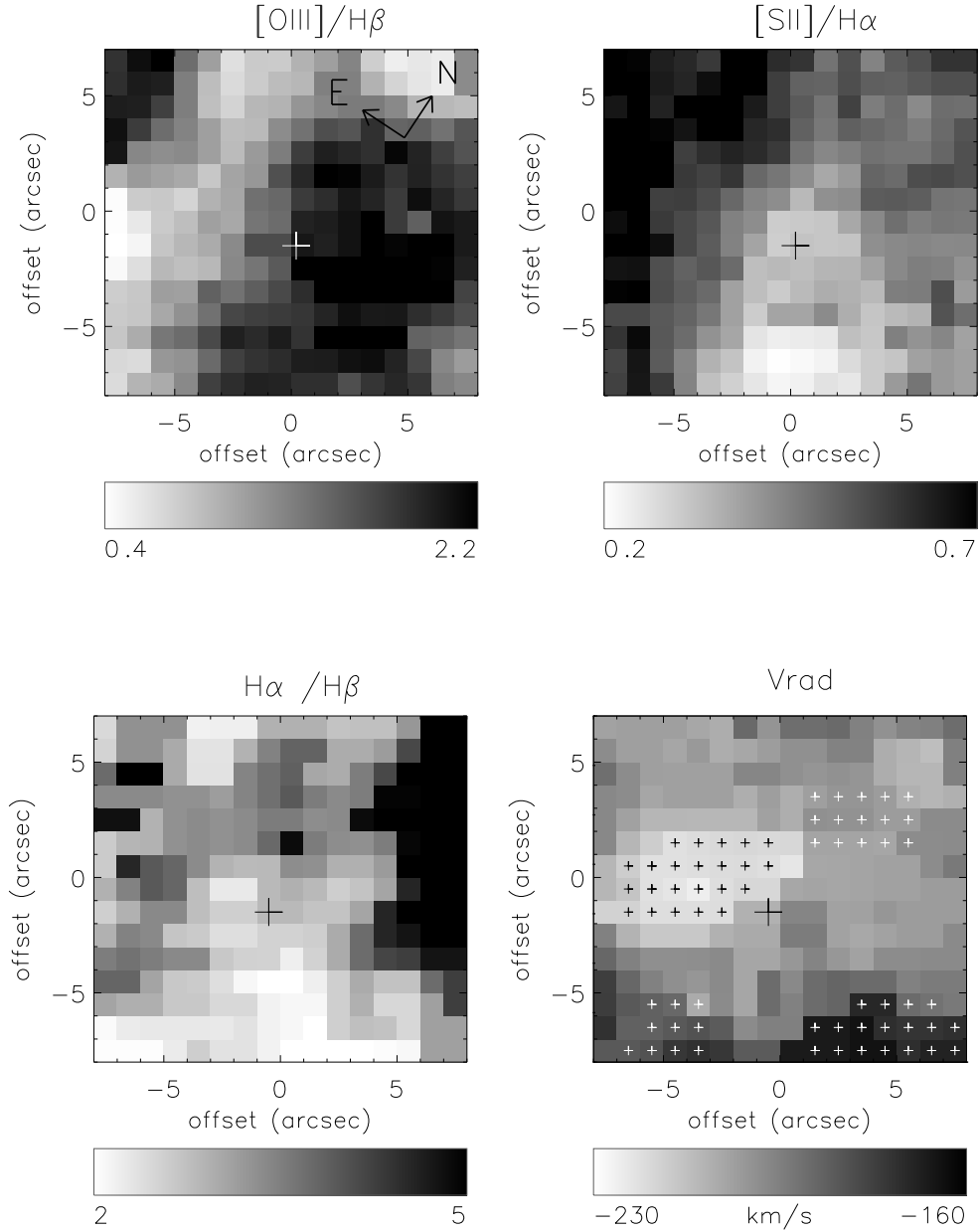}
 \caption{
Ionization maps of B\,416 obtained as ratios of line intensities
of [OIII]~$\lambda 5007$/H$\beta$(narrow),
[SII]~($\lambda 6717+\lambda6731)$/H$\alpha$(narrow),
a map of H$\alpha$/H$\beta$ ratio and
radial velocity map obtained in [SII]~$\lambda 6717$ emission with 
relative flux izophotes of the sum of the lines of the [SII] doublet 
(compare with Fig.\,2). 
Crosses indicate 4 selected regions where mean radial velocities were
measured (see text).
}
\end{figure*}

The lower left panel of Fig.\,6 shows a map of the H$\alpha$/H$\beta$ ratio,
whose canonical value based on recombination theory is expected to be 
$\approx$3. A systematic increase of this value
is seen along the E rim of the ring-like nebula. Most strikingly,
however, the ratio increases to a value of 5 at the N border of the map.
This peculiar region happens to coincide with the aforementioned gap in the
structure of the ring (Fig.\,5), which is thus explained as  
extinction from dust.

A point-by-point analysis of radial velocities in H$\alpha$,
[OIII]\,$\lambda 5007$, and [SII]\,$\lambda 6117$ lines was performed
by fitting gaussians to each line. The internal accuracy of the instrumental
wavelength calibration was increased by measuring differentially against
the [OI]\,$\lambda 5577$ and $\lambda 6300$ night sky lines.
The resulting accuracy of the radial velocity measurements is about 10~km/s
per spatial element (1\sec) in those places where the nebular emission provides
sufficient intensity (Fig.2). The final radial velocity maps were averaged 
over 3 pixels, and the resulting radial velocity accuracy is better than
10~km/s per resolution element.

The lower right panel of Fig.\,6 presents a radial velocity map as derived 
from the deblended [SII]~$\lambda 6717$ emission line. The overplotted 
isophotes were derived from the total intensity of the doublet 
[SII]~$\lambda\lambda 6717,6731$ (cf. Fig.\,2). We chose this 
emission line because it is bright and ubiquitous
over the entire FOV, thus minimizing the uncertainty of our measurement. 
The overall velocity field pattern, however, is the same for all of 
the bright nebular lines, and
shows a systematic variation with a peak-to-peak amplitude of about 
$30-40$~km/s. There are two conspicuous features of extreme radial 
velocities: an area of roughly 9\sec$\times$5\sec across the E rim 
of the nebula has a minimal velocity of $\approx -226$~km/s, 
whereas an area of the same extension and at the same
distance to the W of the star has a significantly more positive 
radial velocity of $\approx -184$~km/s. 
In the N--S direction perpendicular to these two patches the radial 
velocity is constant ($\Delta V_r < 5$~km/s)
and zero with respect to the average, except for the S corner of the field, 
where the velocity is about +10~km/s more positive than the average. 

We derived mean radial velocities of H$\alpha$, [SII]~$\lambda 6717$, 
and [OIII]\,$\lambda 5007$ lines in four selected E, N, W, and S
regions marked by crosses in Fig.\,6.  The mean velocities of 
the H$\alpha$, [SII], and [OIII] lines respectively are 
in the E--region $-239 \pm 3,\, -226 \pm 2,\, -246 \pm 8$~km/s;
in N--region $-226 \pm 2,\, -210 \pm 3,\, -232 \pm 5$~km/s;
in W--region $-212 \pm 3,\, -184 \pm 4,\, -216 \pm 3$~km/s, and
in S--region $-220 \pm 3,\, -197 \pm 7,\, -213 \pm 4$~km/s.
The errors of the mean values depend both on brightness and
homogeneity of a line emission in these selected regions.
Some systematic differences in radial velocities between these three lines
may appear because they were formed under different physical conditions
(unresolved gas clouds and structures).     
The measurements in the selected regions confirm the total radial velocity
gradient observed in the nebula in the E--W direction. It is 
$\Delta V_r(EW) = 33 \pm 8$~km/s over these three lines, and is 
$\Delta V_r(EW) = 34 \pm 4$~km/s in H$\alpha$ + [SII]~$\lambda 6717$,
which are the most similar both in physical conditions and in 
nebula patterns (Fig.\,2).  

The star's radial velocity in H$\alpha$ (the total profile) is 
$-231.1 \pm 3.5$~km/s. It is intermediate between the E and W opposite
features of extreme radial velocities in H$\alpha$ and closer to 
radial velocity of the E region in accordance with location of the star
inside the ring--like nebula (Fig.\,2). 
The stellar radial velocity may be considered as the systemic
velocity of the whole complex (star + nebula).
However the total stellar 
H$\alpha$ line profile may be distorted because of the 
contribution (50\,\%) by the broad stellar component and stellar wind.  

\begin{figure*}
\centering
\includegraphics[width=17cm]{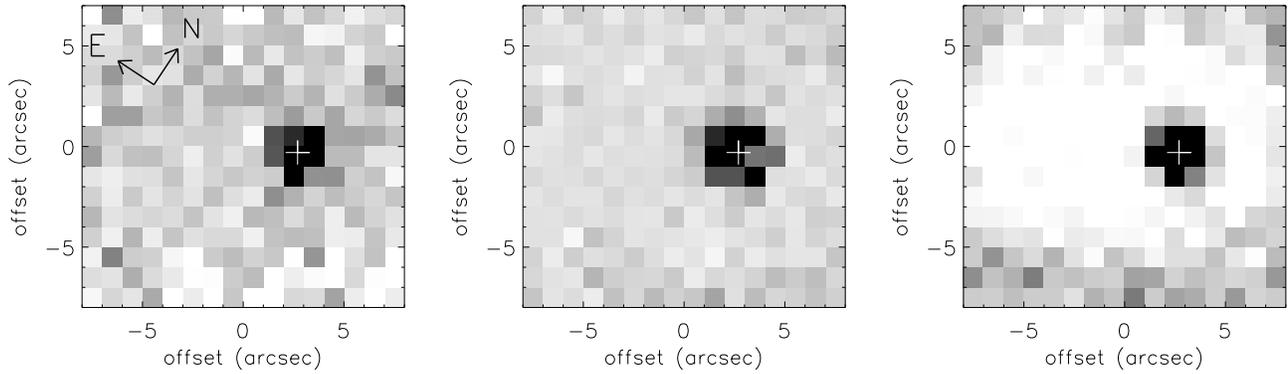}
 \caption{
v\,532 maps in broad H$\beta$ (left), narrow H$\beta$ (middle), and
in the nearby continuum (right panel). Crosses mark the centroid of the star.
}
\end{figure*}

We summarize our findings that the morphological asymmetry of the 
nebula in the E--W direction is accompanied by an asymmetry in the 
radial velocity pattern: the E part of the ring approaches
($\approx -15$~km/s), while the W part receedes ($+15 \div +20$~km/s). 
There is no apparent asymmetry in the N--S direction.
The asymmetry of the nebula in the E--W direction is accompanied also 
by an asymmetry in the gas excitation, while the approaching E part of the 
ring is dominated by low excitation emission (a collisional excitation 
in the rim), but the receeding W part is dominated by high excitation 
gas (a radiative excitation by the central star). 

\subsection*{A nebula around v\,532 (Romano's star)}

Our MPFS images of v\,532, another LBV--candidate in M33,  were taken merely 
in the spectral range 4470~-- 5800~\AA\AA ~(spectral resolution 
$\approx3.5$~\AA).
In the absence of H$\alpha$ and [SII] we had to resort to an analysis 
in H$\beta$, since no [OIII]\,$\lambda 4959,\,5007$
were
detected in the spectrum. Although at first glance the H$\beta$ and continuum
maps of this object appeared point-like, a suspicious slight elongation of the 
stellar image in H$\beta$ with P.A.~$\sim 35^{\circ}$ led us to perform 
a gaussian 
decomposition of the spectral profile into a broad and a narrow component 
along the lines of the exercise with B\,416 as described above. 
As before, we stress that besides an extended nebula, a 
narrow component may either originate in the stellar atmosphere or 
in a nearby unresolved nebula, or in both.
We obtained a good fit from two gaussians with FWHM(narrow)~$=4.1$~\AA~ and 
FWHM(broad)~$=11.1$~\AA.  
   
The resulting maps of v\,532 in broad H$\beta$, narrow H$\beta$, and in the 
nearby continuum are shown in Fig.\,7. Note that the narrow-line component is
a factor of 4 brighter than the broad component. In order to assess the possible
spatial extent of these surface brightness distributions, which are close 
to unresolved point sources, we performed a Moffat fit to the three
maps and inspected their radial intensity profiles for comparison. 
Figure\,8 shows that the narrow component is more extended 
(3\secdot1 FWHM) than the broad component (2\secdot6 FWHM), which in turn 
has the same spatial extent as the continuum intensity profile. 
The difference between the broad and narrow component intensity profiles 
becomes significant at a scale greater than the seeing value, $\approx
2$\sec FWHM. The error bars in the intensity profiles are about
the same. In five consequent bins from 3\sec to 5\sec
the difference between the broad and narrow component energy distributions
varies from 3.5 to 2.5 standard deviations. 
We conclude that an extended nebula has been marginally detected in 
the narrow emission line component of the object.

A radial velocity analysis of the narrow H$\beta$ component supports
this view. We measured the radial velocity in each spaxel differentially against
the night sky [OI]\,$\lambda 5577$ emission line as a reference. The central
part including the star has a velocity $\approx -220$~km/s. The SW part 
of the nebula shows on average positive relative velocities, while 
the velocities of the NE part are mainly negative.
In order to check this finding against any potential errors
from the spectral decomposition, we repeated the velocity measurements
by fitting {\em single} gaussian profiles to the spectra which had been 
coadded from several spaxels in three representative regions 
(Fig.\,9): center (12 spaxels),
NE (9 spaxels), and SW (4 spaxels). This procedure increases the S/N 
ratio in the resulting compound spaxels. 
The derived H$\beta$ radial velocities are $-219 \pm 2.7$~km/s
in the central part, $-191 \pm 10$~km/s in the SW, and $-235 \pm 11$~km/s in
the NE part.

\begin{figure}
\centering
\includegraphics[width=6cm,angle=-90]{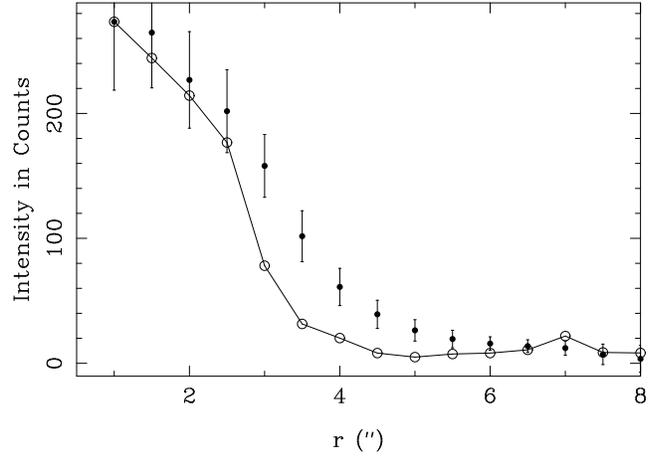}
\vspace{0.5cm}
\caption{
Radial intensity profiles of v\,532 maps in narrow (filled symbols) and broad 
components (open circles) of the H$\beta$ line profile.
}
\end{figure}

\begin{figure}
\centering
\includegraphics[width=7cm]{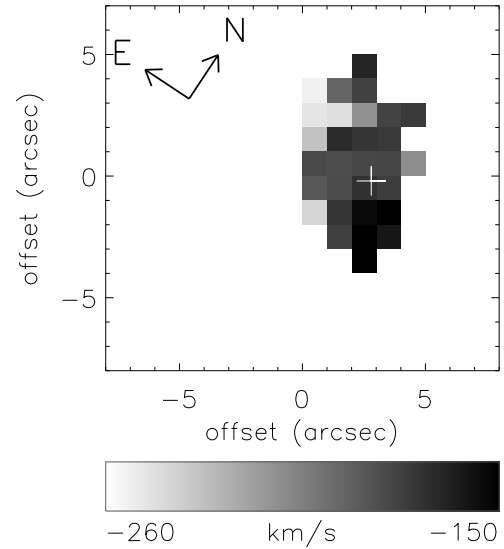}
\vspace{0.5cm}
 \caption{
v\,532 nebula radial velocity map derived in H$\beta$
line. The cross marks the position of the star.
}
\end{figure}

The analysis confirms the presence of a radial velocity gradient across the 
nebula; the SW side of the nebula recedes and the NE side approaches. 
The total radial velocity gradient amounts to $44 \pm 11$~km/s. The 
total angular size of the nebula is $\approx 9\sec$ 
($\approx 30$~pc in projection) in the NE--SW direction.
From the heterogeneous velocity distribution in Fig.\,9 one may suspect that
the system is more complex than just the simple picture of a bipolar 
nebula with receding and approaching lobes. Given the limited spatial 
resolution of our data we therefor forgo any further analysis.

\subsection*{The Spectra of B\,416 and v\,532}

3D and longslit spectra of B\,416 and v\,532 are shown
in Fig.\,10. Three rectified spectra of B\,416 from different instruments 
are plotted with an offset for clarity. The top spectrum was obtained 
with the BTA long-slit spectrograph on 
January 19, 2001 (a spectral resolution FWHM~$\approx 7$\,\AA). 
The middle spectrum
was obtained by coadding spaxels with a digital aperture of the MPFS datacube,
observed on September 28, 1998 (FWHM~$\approx 7$\,\AA). 
The bottom spectrum comes from the INTEGRAL datacube,
which was observed on January 18, 2001 (FWHM~$\approx 11$\,\AA). 
The lower panel in Fig\,10 presents two v\,532 spectra: the top spectrum was
obtained with MPFS on September 18, 1998 (FWHM~$\approx 3.5$\,\AA) 
and the lower spectrum from the long-slit spectrograph,
observed on July 12, 1999 (FWHM~$\approx 3.5$\, \AA). 

\begin{figure*}
\vspace{0.2cm}
\centering
\includegraphics[width=16cm]{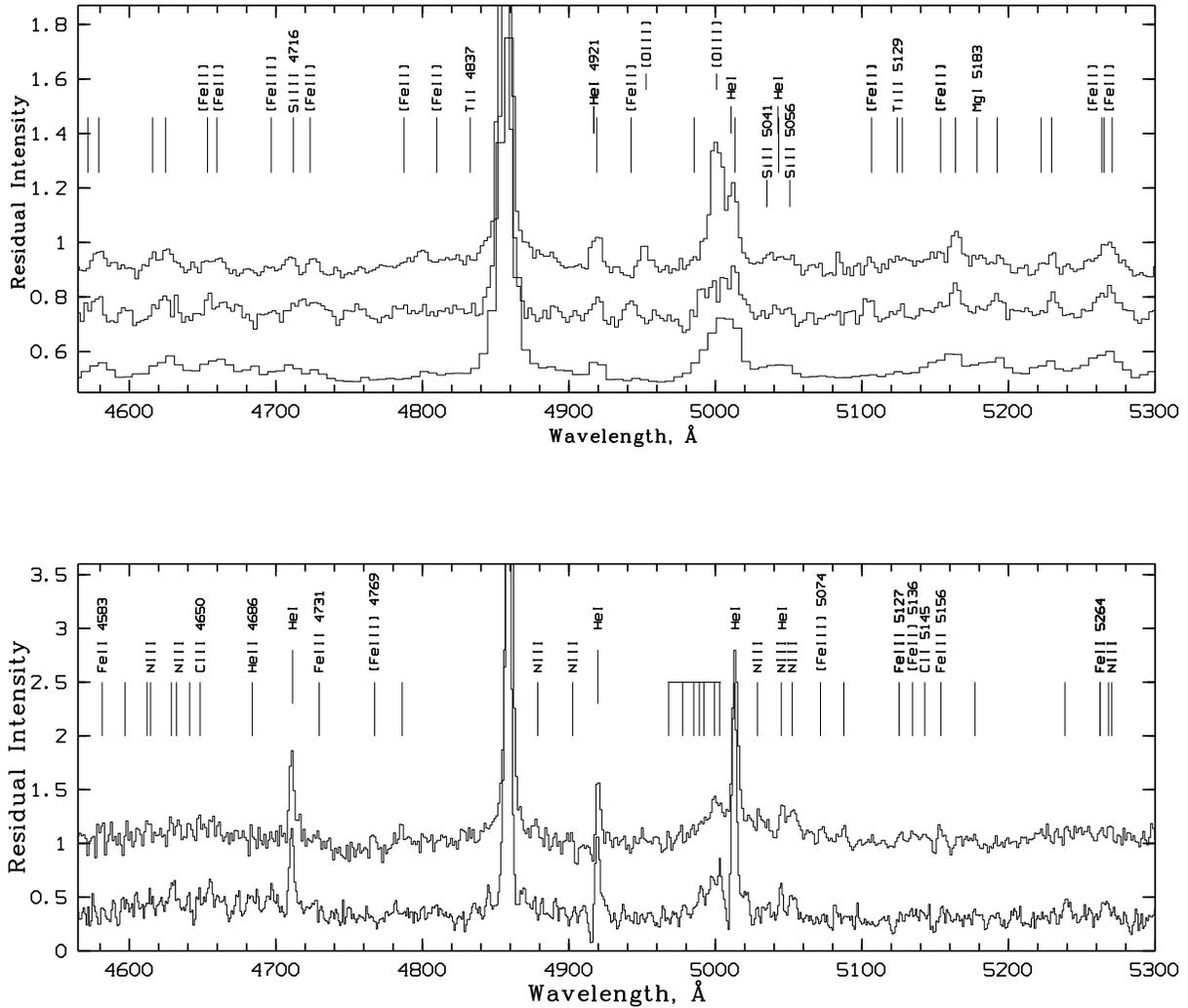}
\vspace*{0.5cm}
 \caption{
Normalized spectra of B\,416 {\bf (top panel)}, and v\,532 
{\bf(bottom panel)}. B\,416 spectra are from INTEGRAL
(bottom), MPFS (middle), and the LS--spectrograph (top).
v\,532 spectra are from the LS--spectrograph (bottom) and MPFS (top).
Numerous FeII lines in the B\,416 spectrum and NII lines in v\,532 spectrum 
are not labeled in order not to confuse the plot.
}
\end{figure*}

A comparison of the long-slit (LS) B\,416 spectrum with its 3D
spectroscopy counterparts (MPFS, INTEGRAL) demonstrates, that the former
is severely affected by nebular contamination, e.g.\ 
[OIII]~$\lambda 4959, 5007$. The strongest line [OIII]~$\lambda 5007$
is superposed on a blend of Fe\,II emission lines in B\,416 spectrum.
The nebular lines, however, are absent in the MPFS and INTEGRAL plots, because
the process of 2-dimensional background correction accurately subtracts the 
local contribution of this light. In general, 
the flux distribution along a slit is not really adequate to estimate 
the background at the location of the object.

We inspected the spectra for spectral variability. According to
Shemmer et al.~(2000), B\,416 shows only weak photometric and spectral
variability, which is more or less in line with our observations taken
essentially at two epochs in 1998 and 2001 (the LS and INTEGRAL spectra were 
coincidently taken only one night apart from each other). 
Considering the difference
in spectral resolution, the LS spectra are indistinguishable from 
INTEGRAL, despite
the systematic [OIII] contamination at $\lambda\lambda 4959, 5007$. 
There are some 
small differences between LS and MPFS spectra.
The situation is quite different in the case of v\,532, which shows a 
pronounced onset of P\,Cyp profiles in He$\,$I between 1998 (MPFS) 
and 1999 (LS).

The most prominent lines in both stars are the strong emission
hydrogen lines with broad wings and narrower He\,I lines.
He\,I~$\lambda 4922, 5015$ lines in B\,416 are blended with the brighter 
Fe\,II~$\lambda 4923, 5018$ lines, but the single He\,I~$\lambda 5876, 
6678$ lines are clearly present in the red part.

The main difference in the spectra of these stars consists, however, in the 
chemical composition of their atmospheres. There 
are many Fe\,II lines in B\,416 and N\,II lines in v\,532. 
In B\,416 besides Fe\,II, Fe\,III,
and He\,I lines, the following ions are also detected:
Si\,II, Cr\,II, N\,II, Ti\,II, S\,II, C\,II, and also neutral elements such
as Ti\,I and Mg\,I are present in the atmosphere.
The Fe\,II wavelengths are indicated in the B\,416 spectrum
(unlabeled in Fig.\,10), while the same is true for some N\,II lines in 
the v\,532 spectrum. 

The star v\,532 exhibits appreciable spectral and photometrical
variability (Fabrika~2000; Kurtev et al.~2001; Polcaro et al.~2003).
In some epochs strong P\,Cyg components appear in He\,I or the hydrogen lines. 
Sometimes the He\,I emission component disappears, sometimes it 
becomes very bright, and the
He\,II~$\lambda\, 4686$ line appears in emission. 
In our MPFS spectrum the He\,I lines are bright with weak blue--shifted
absorption components. A P\,Cyg line profile is more clearly seen in 
He\,I~$\lambda 4922$, split by 220~km/s. 
A faint He\,II~$\lambda\, 4686$ line is possibly present (Fig.\,10) 
with P\,Cyg--like profile. The spectrum is strongly blended by
weak emission lines of N\,II, N\,III, Fe\,II, Fe\,III, C\,II, C\,III,
Si\,II, Si\,III. Faint emissions of Ti, Cr, and S are also observed. 
The most numerical are nitrogen lines. A typical N\,II broad emission 
blend $\approx$ 5000~\AA\ is observed.
The signal--to--noise value in the MPFS spectrum 
is at least 30 in 4700~\AA~ and
5300~\AA\ regions, and is no less than 40 around 5000~\AA.

Both stars possess extended atmospheres, as is
apparent from the presence of the forbidden [FeIII] and [FeII] lines, 
the brightest of which are marked in the spectrum. We have also 
detected the [NII]~$\lambda 5754.8$ forbidden line in both stars 
(not shown in Fig.\,10). 

\section*{Discussion and conclusions}

One may suggest that B\,416 is surrounded by an expanding 
ring--like nebula, whose overall morphological, kinematic, and gas
excitation properties are compatible with an outflow from the star,
which happened to be confined to a plane.
If the ring-like appearance were due to an increased
column-density near the edge of an assumed spheroidal nebula, we would
not expect to find the two patches of extreme radial velocities confined
to the approaching and receding parts of the ring.
It is possible that the plane of the torus is coincident with the
equatorial plane of the star.
The interior region of the nebula is filled with hot, high excitation gas.
The pronounced low excitation emission near the edge of the ring
indicates a shock ionization of material running into the interstellar
medium.

Fitting an ellipse to the entire H$_\alpha$ images of Massey et al. (2001)
in Fig.\,5 and of Shemmer et al. (2000) and assuming an
intrinsic circular symmetry of the ring, we estimate the ring axis
inclination to the line of sight as $i \sim 30^{\circ}$.
From the apparent radial velocity gradient of $\approx 20$~km/s,
we roughly estimate the deprojected physical expansion velocity
as $\sim 40$~km/s. The physical dimension of the nebula is measured from
the major axis along the N--S direction as a radius of $\approx~30$~pc
(9\sec). From these numbers we find a dynamical lifetime of the nebula of
$t_d \sim 8 \cdot 10^5$~years. 
This estimate indicates a main--sequence or a pre--LBV
status of the star. The nebula is probably a residual MS bubble.
Using the average density of the nebula found above ($n_e \sim
10\,cm^{-3}$), 
we estimate that a total mass of the nebula is about a few thousand
solar masses. Such a big nebula may be composed mainly of swept up
interstellar gas.

The photometry of this star has been studied well (Shemmer et al.~2000),
but it does not show the variability that is typical of LBVs.
Using the Sternberg Institute plate collection, Zharova (2004)
has found that for the last 50 years a variability of B\,416 was notably
less than 1 magnitude. There are instead minor long--term (1987~-- 1997)
variations of about 0\magdot15 (Shemmer et al.~2000).  

Spectrally, B\,416 has to be classified as an `iron--star'' (Walborn and
Fitzpatrick~2000). It is the brightest supergiant in M33 with UV--excess
(Massey et al. 1996). Based on spectral criteria, luminosity,
and the absence of a large--amplitude variability, B\,416 may be
classified as B[e]--supergiant star (Zickgraf et al.~1986;
Lamers et al.,~1998b). The spectrum is typical of
[Be]--supergiants (or `sg[Be]''stars,
Lamers et al.~1998b)~--- very bright hydrogen lines, a forest of
narrower Fe\,II lines, He\,I lines, forbidden lines.

The presence of anisotropical mass loss is a common and distinguishing 
property of B[e]--supergiants (Lamers et al.,~1998b). UV resonance
line profiles in the edge--on inclination stars show that the stellar winds
are very slow in massive B[e]--supergiants (Zickgraf~2003). The expansion 
velocities are $\sim 70-100$~km/s, a factor of 10 less than usually 
observed in stars of a similar spectral type. Such velocities are also typical
of disk--like outflows in close binaries, where a donor star overfills 
its critical Roche lobe. Binaries among B[e]--supergiants were reviewed
by Zickgraf (2003), 
while Sholukhova et al. (2004) have studied radial velocities
in B\,416 and found the periodical variability with a period $\sim 16$
days. We postpone a discussion of B\,416 as a close binary for a future paper.

The second star of our study, v\,532 shows a typical LBV--like variability
(Romano~1978; Kurtev et al.~2001; Sholukhova et al.~2002).
Over the past 50 years it has changed its brightness over a range of
$B= 16.3 \div 18.4$.
There are large amplitude variations with a time--scale of years, onto which
a sporadic 0\magdot3~-- 0\magdot5 amplitude variability is superimposed with
a time--scale of several months. 
The last irregular minimum occurred from 1974 to 1981, and the last 
(also quite irregular) maximum was observed on 1990~-- 1995. At the time 
of our spectroscopic observations
in September 1998 the star had intermediate, gradually decreasing brightness.
We estimate the brightness of v\,532 
during this epoch as $B= 17\magdot5 \pm 0\magdot15$ (Sholukhova et al.~2002).

During an intermediate brightness stage in 1998 -- 2001, the star went 
into a higher excitation Ofpe/WN9--like spectrum (Fig.~10). 
The spectrum is closest to 
those of WN\,10-11 stars, which were introduced by Smith et al.~(1994) and 
studied extensively in other papers (Crowther et al.~1995a\,b\,c;
Crowther and Smith~1997; Bohannan and Crowther~1999).
The most similar that we could identify in the literature are
the well--known LBV AG\,Car during minimum (Hutsemehers and Kohoutek~1988, 
Stahl~1986, Smith et al.~1994), He\,3--519 (Smith et al.~1994), and 
S\,142 (Crowther and Smith~1997). The major criteria for a very 
late WN\,10-11 classification (Crowther and Smith~1997) 
are confirmed in the spectrum of v\,532:
i) a rich low excitation nitrogen spectrum, where N\,II lines are brighter than 
those of N\,III; ii) narrow He\,I lines, faint He\,II~$\lambda 4686$;
iii)  forbidden [N\,II]~$\lambda 5755$ and [Fe\,III] lines; iv) 
the expansion velocities derived from He\,I P\,Cyg--like profiles are low,
$v_{\infty}=100 \div 300$~km/s (it is 220~km/s in v\,532); 
v) the presence of a nebula with a low expansion velocity on the order 
of a few tens of km/s (marginally detected in this study). 

The spectrum of v\,532 is similar to the spectra of Ofpe/WN9 stars.
These stars are believed to be directly related to LBVs as in fact,
``dormant'' LBVs. Transitions of LBV\,$\leftrightarrow$\,Ofpe/WN9 have been
observed and are well--known, for example in AG\,Car, or R\,127
(Stahl et al.~1983; Stahl~1986). We saw the same transition in
v\,532, when during the peak of the maximum phase (fall 1992) the star
presented a low excitation LBV--like spectrum (Szeifert~1996; Fabrika~2000).

The star v\,532 thus has to be classified thus as an LBV object based on 
its spectrum, spectral, and photometrical variability. 
In its current state the star may join the
group of WN11 stars. The nitrogen spectrum of v\,532
indicates an evolved status, where nitrogen produced in the
core appears at the surface. 

The B\,416 spectrum and its nebula age 
($\sim 8 \cdot 10^5$~years) indicate an MS or pre--LBV status 
(hydrogen--burning) of the star (Schaerer et al.~1993). The disk--like 
geometry of the nebula around B\,416 does not contradict
the main sequence status of B\,416. An asymmetry in the mass loss 
might be due to fast rotation or close binarity. 

\begin{acknowledgements}

The authors thank S.N.\,Dodonov and A.N.\,Burenkov
for help during the observations. S.Fabrika and O. Sholukhova
are grateful to the AIP for hospitality. 
This work has been supported by the RFBR grants N\,03--02--16341,
N\,04--02--16349. O. Sholukhova acknowledges support by INTAS grant 
YSF 2002-281.

\end{acknowledgements}


\begin{thebibliography}{99}
\bibitem{}
Afanasiev, V.L. 1998, (html://www.sao.ru/~gafan/devices/mpfs/mpfs\_main.htm)

\bibitem[Arribas et al. (1998)] {arribas1998b} Arribas, S., Cavaller, L.,
   Garcia-Lorenzo, B., Garcia-Marin, A., Herreros, J.M., Mediavilla, E.,
   Pi, M., del Burgo, C., Fuentes, J., Rasilla, J.L., Sosa, N. 1998b,
   in {\it Fiber Optics in Astronomy III}, eds.\ S. Arribas, E. Mediavilla, 
   F. Watson, ASP Conf. Ser. Vol.152, p. 149

\bibitem{}
Artyukhina, N.M., Goranskii, V.P., Gorynya, N.A. et al. 1995, General Cataloque
of variable stars. Vol 5, Moscow, Kosmosinform.

\bibitem[Becker 2002] {becker2002} Becker, T. 2002, Thesis, 
   University of Potsdam

\bibitem[Becker et al.\ 2003] {becker2003} Becker, T., Fabrika, S., Roth, M.M.
   2004, Astronomische Nachrichten, 325, 155

\bibitem{}
Bohannan, B. \& Crowther, P.A. 1999, ApJ, 511, 374

\bibitem{}
Calzetti, D., Kinney, A.L., Ford, H., Dogget, J. \& Long, K.S. 1995, AJ, 110, 2739

\bibitem{}
Conti, P.S. 1976, in Proc. IAU Symp. 70 "Be and shell stars", Sletteback A.
ed., (Dordrecht: Reidel), 447

\bibitem{}
Corral, L. 1996, AJ, 112, 1450

\bibitem{}
Crowther, P.A., Hiller, D.J. \& Smith, L.J. 1995a, A\&A, 293, 172

\bibitem{}
Crowther, P.A., Hiller, D.J. \& Smith, L.J. 1995b, A\&A, 293, 403

\bibitem{}
Crowther, P.A., Smith, L.J., Hiller, D.J. \& Schmutz, W. 1995c, A\&A, 293, 427

\bibitem{}
Crowther, P.A. \& Smith, L.J. 1997, A\&A, 320, 500


\bibitem{}
Chu, You-Hua, Chen, C.-H. Rosie, Danforth, C. et al. 2003, Astron. J.,
125, 2098

\bibitem{}
Fabrika, S. \& Sholukhova, O. 1995, Ap\&SS, 226, 229

\bibitem{}
Fabrika, S. 2000, in ASP Conf. Ser. 204, Thermal and ionization aspects of flows from hot stars:
observations and theory, eds H.J.G.L.M. Lamers and A. Sapar,
57.

\bibitem{}
Fabrika, S. \& Sholukhova, O. 1999, A\&AS, 140, 309

\bibitem{}
Figer, D.F., Morris, M., Geballe, T.R., Rich, R.M., Serabyn, E.,
McLean, I.S., Puetter, R.C. \& Yahil, A. 1999, ApJ, 525, 759

\bibitem{}
Garcia-Segura, G., Mac Low, M. \& Langer, N., 1996, A\&A, 305, 229

\bibitem{}
Humphreys, R.M. \& Davidson, K. 1994, PASP, 106, 1025

\bibitem{}
Humphreys, R.M. \& Sandage, A. 1980, ApJS, 44, 319

\bibitem{}
Hutsemehers, D. \& Kohoutek, L. 1988, A\&AS, 73, 217

\bibitem{}
Kudritzki, R.P. 1998, in Proceedings of the 8th Canary
Winter School, Stellar Astrophysics for the Local Group. A first step
to the Universe, eds. A.\,Aparicio, A.\,Herrero, F\,Sanchez,
New York, Cambridhe Univ. Press.,149

\bibitem{}
Kurtev, R., Sholukhova, O., Borissova, J. \& Georgiev, L. 2001, Rev. Mex.
Astron. Astrofis., 37, 57

\bibitem{}
Lamers, H.J.G.L.M., Bastiaanse, M.V., Aerts, C. \& Spoon, H.W.W.
1998a, A\&A, 335, 605

\bibitem{}
Lamers, H.J.G.L.M., Zickgraf, F.-I., de Winter, D., Houziaux, L. \&
Zorec, J. 1998b, A\&A, 340, 117

\bibitem{}
Lamers, H.J.G.L.M., Nota, A., Panagia, N., Smith, L.J. \& Langer, N. 2001,
ApJ, 551, 764

\bibitem{}
Massey, P., Bianchi, L., Hutchings, J.B. \& Stecher, T.P. 1996, ApJ, 469, 647

\bibitem{}
Massey, P., Hodge, P. W., Holmes, S., Jacoby, G., King, N. L., Olsen, K., 
Saha, A. \& Smith, C. 2001,  AAS 199th meeting, BAAS, 33, 1496 

\bibitem{}
Nota, A. 1999, in Variable and Non--spherical Stellar Winds in
Luminous Hot Stars, eds. B. Wolf, O. Stahl, A.W. Fullerton, Springer, 62

\bibitem[Osterbrock1989]{osterbrock1989} Osterbrock, D.E. 1989,
   {\em Astrophysics of Gaseous Nebulae and Active Galactic Nuclei},
   University Science Books, Mill Valley

\bibitem{}
Polcaro, V.F. Gualandi, R. Norci, L. Rossi, C., Viotti, R.F. 2003,
A\&A, 411, 193

\bibitem{}
Romano, G. 1978, A\&A, 67, 291

\bibitem[Roth et al.(2000)]{roth2000} Roth, M.~M.~et al.\ 2000, 
   SPIE, 4008, 277 

\bibitem{} Roth, M.M., Becker, T., Kelz, A., Schmoll, J. 2003,
   astro-ph/0311407

\bibitem{}
Sabbadin, F. \& D'Odorico, S.D. 1976, A\&A, 49, 119

\bibitem{}
Schaerer, D., Meynet, G., Maeder, A. \& Schaller, G. 1993, A\&AS, 98, 523

\bibitem{}
Shemmer, O. \& Leibowitz, E.M. 1998, IBVS, N\,4595

\bibitem{}
Shemmer, O., Leibowitz, E.M. \& Szkody, P. 2000, MNRAS, 311, 698

\bibitem{}
Sholukhova, O.N., Fabrika, S.N., Vlasyuk, V.V. \& Burenkov, A.N. 1997,
Astron. Letters, 23, 458

\bibitem{}
Sholukhova, O.N., Fabrika, S.N. \& Vlasyuk, V.V. 1999,
Astron. Letters, 25, 14

\bibitem{}
Sholukhova, O., Zharova, A., Fabrika, S. et al., 2002,
Radial and Nonradial Pulsations as Probes of Stellar Physics,
ASP Conf. Proceedings, 259. Edited by Conny Aerts,
Timothy R. Bedding, and Jorgen Christensen-Dalsgaard, 522

\bibitem{}                                                       
Sholukhova, O., Fabrika, S., Roth, M. \& Becker, 2004,           
Baltic. Astron., 13, 156 

\bibitem{}
Smith, R.C., Kirshner, R.P., Blair, W.P., Long, K.S. \&
Winkler, P.F. 1993, ApJ, 407, 564

\bibitem{}
Smith, L.J., Crowther, P.A. \& Prinja, R.K. 1994, A\&A, 281, 833

\bibitem{}
Stahl, O. 1986, A\&A, 164, 321

\bibitem{}
Stahl, O., Wolf, B., Klare, G., Cassatella, A., Krauter, J.,
  Persi, P. \& Ferrari-Toniolo, M. 1983, A\&A, 127, 49

\bibitem{}
Szeifert, T. 1996, in $33^{rd}$ Liege Institute Astroph. Coll., Wolf--Rayet
stars in the framework of stellar evolution, eds. J.M.Vreux, A.Detal,
D.Fraipont-Caro, E.Gosset, and G.Rauw, (Liege), 459

\bibitem{}
Walborn, N.R. \& Fitzpatrick, E.L. 2000, PASP, 112, 50

\bibitem{}
Weis, K. 2003, A\&A, 408, 205   

\bibitem{}
Zharova, A. 2004, private communication

\bibitem{}
Zickgraf, F.-J., Wolf, B., Stahl, O. \& Appenzeller, I. 1986, A\&A, 163, 119

\bibitem{}
Zickgraf, F.-J., Wolf, B., Stahl, O., Leitherer, C. \& Klare, G. 1985, A\&A, 143, 421

\bibitem{}
Zickgraf, F-J. 2003, A\&A, 408, 257

\end{thebibliography}
\end{document}